\DeclareFontFamily{OT1}{msb}{}{}
\DeclareFontShape{OT1}{msb}{m}{n}
 {  <5> <6> <7> <8> <9> <10> gen * msbm
      <10.95><12><14.4><17.28><20.74><24.88>msbm10}{}
\DeclareMathAlphabet{\bubble}{OT1}{msb}{m}{n}

\def\bZ{{\bubble Z}}
\def\bN{{\bubble N}}

\documentstyle[12pt]{article}
\parskip=\medskipamount
\begin{document}

\def\l#1#2{\raisebox{.0ex}{$\displaystyle
  \mathop{#1}^{{\scriptstyle #2}\rightarrow}$}}
\def\r#1#2{\raisebox{.0ex}{$\displaystyle
\mathop{#1}^{\leftarrow {\scriptstyle #2}}$}}

\newcommand{\p}[1]{(\ref{#1})}
\newcommand{\sect}[1]{\setcounter{equation}{0}\section{#1}}

\makeatletter
\def\eqnarray{\stepcounter{equation}\let\@currentlabel=\theequation
\global\@eqnswtrue
\global\@eqcnt\z@\tabskip\@centering\let\\=\@eqncr
$$\halign to \displaywidth\bgroup\@eqnsel\hskip\@centering
  $\displaystyle\tabskip\z@{##}$&\global\@eqcnt\@ne
  \hfil$\displaystyle{\hbox{}##\hbox{}}$\hfil
  &\global\@eqcnt\tw@ $\displaystyle\tabskip\z@
  {##}$\hfil\tabskip\@centering&\llap{##}\tabskip\z@\cr}
\@addtoreset{equation}{section}
\makeatother


\renewcommand{\thefootnote}{\fnsymbol{footnote}}
\newpage
\setcounter{page}{0}
\pagestyle{empty}
\begin{flushright}
{January 2004}\\
{ITP-UH-03/04}\\
{nlin.SI/0401029}
\end{flushright}
\vfill

\begin{center}
{\LARGE {\bf A note on fermionic flows of the }}\\[0.3cm]
{\LARGE {\bf N=(1$|$1) supersymmetric Toda lattice hierarchy
}}\\[1cm]

{\large Olaf Lechtenfeld$^{a,1}$ and Alexander S. Sorin$^{b,2}$}
{}~\\
\quad \\
{\em {~$~^{(a)}$ Institut f\"ur Theoretische Physik, Universit\"at
Hannover,}}\\
{\em Appelstra\ss{}e 2, D-30167 Hannover, Germany}\\[10pt]
{\em {~$~^{(b)}$ Bogoliubov Laboratory of Theoretical Physics, JINR,}}\\
{\em 141980 Dubna, Moscow Region, Russia}~\quad\\

\end{center}

\vfill

\centerline{{\bf Abstract}}
\noindent
We extend the Sato equations of the 
$N{=}(1|1)$ supersymmetric Toda lattice hierarchy by 
two new infinite series of fermionic flows 
and demonstrate that the algebra of the flows
of the extended hierarchy is 
the Borel subalgebra of the 
$N{=}(2|2)$ loop superalgebra.

{}~

{\it PACS}: 02.20.Sv; 02.30.Jr; 11.30.Pb

{\it Keywords}: Completely integrable systems; Toda field theory;
Supersymmetry; Discrete symmetries

\vfill
{\em E-Mail:\\
1) lechtenf@itp.uni-hannover.de\\
2) sorin@thsun1.jinr.ru }
\newpage
\pagestyle{plain}
\renewcommand{\thefootnote}{\arabic{footnote}}
\setcounter{footnote}{0}

In this note, we consider the integrable $N{=}(1|1)$ supersymmetric
generalization~\cite{3} of the two-dimensional bosonic Toda lattice 
hierarchy (2DTL hierarchy)~\cite{2}. It is given by an infinite system 
of evolution equations (flows) for an infinite set of bosonic and 
fermionic lattice fields evolving in two bosonic and two fermionic
infinite ``towers" of times. A subsystem of the 2DTL hierarchy involves an 
$N{=}(1|1)$ supersymmetric integrable generalization of the 2DTL equation, 
which is called the $N{=}(1|1)$ 2DTL equation.

Two new infinite series of fermionic flows of the $N{=}(1|1)$ 2DTL hierarchy
were constructed in~\cite{6} in a heuristic way by solving symmetry equations
corresponding to the $N{=}(1|1)$ 2DTL derived in~\cite{5}.
This hierarchy was shown to actually have a higher symmetry, namely an 
$N=(2|2)$ supersymmetry~\cite{8,7}. Together with the previously known 
bosonic and fermionic flows of the $N{=}(1|1)$ 2DTL hierarchy, these flows 
are symmetries of the $N{=}(1|1)$ 2DTL equation.

The existence of those additional fermionic symmetries implies that 
the Lax pair or Sato equations proposed in~\cite{3} are incomplete because 
they lack the fermionic flows corresponding to these symmetries. Therefore, 
the following task arises: Can one construct a complete description of this 
hierarchy including the additional fermionic flows? Let us mention that 
a similar problem was partly discussed in~\cite{4} in a slightly different 
context.   

This brief note addresses the above question. We extend the Sato equations 
of the paper~\cite{3} by new equations which indeed describe the additional
fermionic flows constructed in~\cite{6}. With these, the complete algebra 
of flows of the extended hierarchy is confirmed. 

Our starting point is the Sato equations of the 
$N{=}(1|1)$ 2DTL hierarchy ~\cite{3} 
\begin{eqnarray}
&&D_n^{\pm}W^{\mp}=\Big(({L^{\pm}})^n_{*}\Big)_{\pm}
W^{\mp}-{W^{\mp}}^{{*}(n)}(\Lambda^{\pm})^n, \nonumber\\
&&D^{\pm}_n W^{\pm}=\Big(({L^{\pm}})^n_{*}\Big)_{\pm}W^{\pm},
\quad n \in {\bN} ,
\label{1}
\end{eqnarray}
where Lax operators 
$L^{\pm}$ and dressing operators $W^{\mp}$ are given by
\begin{eqnarray}
L^{\pm}=(W^{\mp})^{*}\Lambda^{\pm}(W^{\mp})^{-1}, \quad
\Lambda^{\pm}\equiv (\pm 1)^{j+1}e^{\pm\partial} 
\label{5a}
\end{eqnarray}
and
\begin{eqnarray}
W^{\pm}=\sum_{k=0}^{\infty}
w^{\pm}_{k,j}
e^{\pm k\partial}, \quad
w^{-}_{0,j}=1,
\label{4}
\end{eqnarray}
respectively. Here, 
$w^{\pm}_{2k,j}$ 
($w^{\pm}_{2k+1,j}$) are bosonic (fermionic) lattice
fields ($j \in {\bZ}$). 
The operator
$e^{l{\partial}}$ ($l \in {\bZ}$) 
is the discrete lattice shift which acts
according to the rule
\begin{eqnarray}
e^{l{\partial}}w^{\pm}_{k,j}\equiv w^{\pm}_{k,j+l} e^{l{\partial}},
\label{rule1}
\end{eqnarray}
and the subscript $+$ ($-$) distinguishes the part of an operator 
which includes operators $e^{l{\partial}}$ at $l\geq 0$ ($l< 0$). 
The symbols $D^{\pm}_{2n}$ and $D^{\pm}_{2n+1}$  denote bosonic 
and fermionic evolution derivatives, respectively. 
The subscripts $*$ and superscripts $*{(n)}$ and $*$ 
are defined according to the rules
\begin{eqnarray}
&&(L^{\pm})^{2n}_{*}:=((L^{\pm})^{*}L^{\pm})^{n}, \quad
(L^{\pm})^{2n+1}_{*}:=L^{\pm}((L^{\pm})^{*}L^{\pm})^{n}, \nonumber\\
&&(W^{\pm})^{*(2n)}:=W^{\pm}, \quad
(W^{\pm})^{*(2n+1)}:=(W^{\pm})^{*}, \nonumber\\
&& (W^{\pm}[w^{\pm}_{k,j}])^{*}:=W^{\pm}[{w^{\pm}_{k,j}}^{*}],
\quad (w^{\pm}_{k,j})^{*}:=(-1)^k w^{\pm}_{k,j}.
\label{rule2}
\end{eqnarray}

The flows 
\p{1} generates the Borel subalgebra of the 
$N{=}(1|1)$ loop superalgebra ~\cite{3} 
\begin{eqnarray}
\left[ D_{2n}^{\pm} , D_k^{\pm} \right]=
\left[D^+_n , D^-_k \right\}=0, \quad
\left\{ D^{\pm}_{2k+1} , D^{\pm}_{2l+1}
\right\}=2D^{\pm}_{2(k+l+1)}. 
\label{8a}
\end{eqnarray}

The $N=(1|1)$ supersymmetric 2DTL equation belongs to the system of
equations \p{1}. Indeed, using 
\p{1} 
one can easily derive the
following equations
\begin{eqnarray}
D^{+}_1w^+_{0,j}&=&-(w^-_{1,j}+w^-_{1,j+1})w^+_{0,j}, \nonumber\\
D^{-}_1 w^-_{1,j}&=&(-1)^{j+1}\frac{w^+_{0,j}}{w^+_{0,j-1}}.
\label{eqs}
\end{eqnarray}
Then, eliminating the field $w^-_{1,j}$ from \p{eqs} and introducing
the notation 
\begin{eqnarray}
v_{0,j}:=(-1)^{i+1}\frac{w^+_{0,j}}{w^+_{0,j-1}} 
\label{11}
\end{eqnarray}
we obtain the equation
\begin{eqnarray}
D^{+}_1D^{-}_1 \ln v_{0,j}= v_{0,j+1} - v_{0,j-1}
\label{toda}
\end{eqnarray}
which reproduces the $N=(1|1)$ superfield form of the
$N=(1|1)$ 2DTL equation.

We propose to extend consistently the flows \p{1} by
the following two new infinite series of fermionic flows, 
\begin{eqnarray}
&&\widehat{{\cal D}}^{\pm}_{2k+1}W^{\mp}=\Big(({\widehat
L^{\pm}})^{2k+1}_{*}\Big)_{\pm}
W^{\mp}-{W^{\mp}}^{*}({\Pi^{\pm})}^{2k+1}, \nonumber\\
&&\widehat{{\cal D}}^{\pm}_{2k+1}W^{\pm}=
\Big(({\widehat L^{\pm}})^{2k+1}_{*}\Big)_{\pm} W^{\pm},
\quad k \in {\bN} ,
\label{2}
\end{eqnarray}
where
\begin{eqnarray}
\widehat{L}^{\pm}=(W^{\mp})^{*}\Pi^{\pm}(W^{\mp})^{-1},
\quad \Pi^{\pm}\equiv (\mp1)^j e^{\pm\partial}
\label{5}
\end{eqnarray}
and $\widehat{{\cal D}}^{\pm}_{2k+1}$ 
denote new fermionic evolution derivatives.
The calculation of their algebra yields
\begin{eqnarray}
&&\left\{\widehat{{\cal D}}^{\pm}_{2k+1} , \widehat{{\cal
D}}^{\mp}_{2l+1}\right\}=0, \quad
\left\{ \widehat{{\cal D}}^{\pm}_{2k+1} , \widehat{{\cal D}}^{\pm}_{2l+1}  
\right\}= 2(-1)^{k+l+1}D^{\pm}_{2(k+l+1)}, \nonumber\\
&&\left[ \widehat{{\cal D}}^{\pm}_{2k+1} , D^{\pm}_l \right\}=
\left[ \widehat{{\cal D}}^{\pm}_{2k+1} , D^{\mp}_l \right\}=0,
\label{8}
\end{eqnarray}
which, together with \p{8a}, form the Borel subalgebra of the $N{=}(2|2)$ 
loop superalgebra. Hence, the extended supersymmetric hierarchy with the 
flows \p{1} and \p{2} can be called the $N{=}(2|2)$ supersymmetric 2DTL
hierarchy. Thus, we have finally established the origin of the
fermionic symmetries observed in \cite{6}: they are exactly the flows 
\p{2} of the extended hierarchy.

{}~

\noindent{\bf Acknowledgments.}
This work was partially supported by RFBR Grant No.
03-01-00781, RFBR-DFG Grant No. 02-02-04002,
DFG Grant 436 RUS 113/669/0-2 and by the Heisenberg-Landau program.

{}~


\end{document}